\documentclass[preprint,12pt]{elsarticle}
\usepackage{amssymb}
\usepackage{mathrsfs}
\usepackage{epic,eepic,epsf,epsfig}
\usepackage{graphicx}

\usepackage{amsmath}
\usepackage{amsbsy}
\usepackage{amsfonts}

\newtheorem{theorem}{Theorem}[section]%
\newtheorem{lem}[theorem]{Lemma}%

\def\f{\noindent}

\journal{Information Processing Letters}

\begin{document}

\begin{frontmatter}

\title{$3$-extra connectivity of $3$-ary $n$-cube networks}

\author{Meimei Gu, Rongxia Hao$^{\ast}$\\
{\small\em Department of Mathematics, Beijing Jiaotong University, 100044, China}}


\fntext[1]{$^{\ast}$Corresponding author}
\fntext[2]{E-mail addresses: 12121620$@$bjtu.edu.cn (Meimei Gu),
 rxhao$@$bjtu.edu.cn (Rongxia Hao)}

\begin{abstract}
Let $G$ be a connected graph and $S$ be a set of vertices. The $h$-extra connectivity of $G$ is the cardinality of a minimum set $S$ such that $G-S$ is disconnected and each component of $G-S$ has at least $h+1$ vertices. The $h$-extra connectivity is an important parameter to measure the reliability and fault tolerance ability of large interconnection networks. The $h$-extra connectivity for $h=1,2$ of $k$-ary $n$-cube are gotten by Hsieh et al. in [Theoretical Computer Science, 443 (2012) 63-69] for $k\geq 4$ and Zhu et al. in [Theory of Computing Systems, arxiv.org/pdf/1105.0991v1 [cs.DM] 5 May 2011] for $k=3$. In this paper, we show that the $h$-extra connectivity of the $3$-ary $n$-cube networks for $h=3$ is equal to $8n-12$, where $n\geq 3$.

\end{abstract}


\end{frontmatter}

\section{Introduction}
It is well known that a topological structure of an interconnection network can be modeled by a loopless undirected graph $G=(V,E)$, where the vertex set $V$ represents the processors and the edge set $E$ represents the communication links. In this paper, we use graphs and networks interchangeably.

Let $G$ be a simple undirected graph. Two vertices $v_{1},v_{2}$ in $V(G)$ are said to be $adjacent$ if and only if $(v_{1},v_{2})\in E(G)$. The $neighborhood$ of a vertex $u$ in $G$ is the set of all vertices adjacent to $u$ in $V(G)$, denoted by $N_{G}(u)$. The cardinality $|N_{G}(u)|$ represents the $degree$ of $u$ in $G$, denoted by $d_{G}(u)$(or simply $d(u)$), $\delta(G)$ the $minimum$ $degree$ of $G$. For a vertex subset $S\subseteq V(G)$, the neighborhood of $S$ in $G$ is $N_{G}(S)=(\bigcup_{u\in S}N_{G}(u))-S$. A $subgraph$ of $G=(V,E)$ is a graph $H=(V^{'},E^{'})$ such that $V^{'}\subseteq V$ and $E^{'}\subseteq E$. For a subgraph $H$ of $G$, $N_{G}(V(H))$ can be simplified as $N_{G}(H)$. For a subset $S$ of $V(G)$, the $induced$ $subgraph$ of $S$, written by $G[S]$, is a subgraph of $G$, whose vertex set is $S$ and an edge $e\in G[S]$ if and only if both end vertices of $e$ are in $S$. $N[S]$ is also used to denote the induced subgraph of $N_{G}(S)\bigcup S$. A subset $S\subseteq V(G)$ is a vertex cut if $G-S$ is disconnected. The components of $G$ are its maximal connected subgraphs.

A $path$ $P_{k}=(v_{1},v_{2},\cdots,v_{k})$ for $k\geq2$ in a graph $G$ is a sequence of distinct vertices such that any two consecutive vertices are adjacent, and $v_{1}$ and $v_{k}$ are the $end$-$vertices$ of the path. For convenience, use $P_{t}$ to denote a path of $t$ vertices. A path of $G$ of length $n$ will be called an $n$-$path$. A $cycle$ $C_{k}=(v_{1},v_{2},\cdots,v_{k},v_{1})$ for $k\geq 3$ is a sequence of vertices in which any two consecutive vertices are adjacent, where $v_{1},v_{2},\cdots,v_{k}$ are all distinct. A cycle of $G$ of length $n$ will be called an $n$-$cycle$. A $complete$ $graph$ of $n$ vertices, denoted by $K_{n}$, is a simple graph whose vertices are pairwise adjacent.

Let $G$ and $H$ be two graphs. $G$ and $H$ are $distinct$ if their vertex sets are different, and $disjoint$ if they have no common vertices. An $isomorphism$ from a graph $G$ to a graph $H$ is a bijection function $\pi: V(G) \rightarrow V(H)$ such that $(u, v)\in E(G)$ if and only if $(\pi(u), \pi(v))\in E(H)$. We write $G\cong H$ if there is an isomorphism from $G$ to $H$.

The $connectivity$ $\kappa(G)$ of a connected graph $G$ is the minimum number of vertices removed to get the graph disconnected or trivial. A graph $G$ is said to be $super$ $connected$, or simply super-$\kappa$, if every minimum vertex cut creates exactly two components, one of which is a singleton. Connectivity as a measure of reliability underestimates the fault tolerance ability of these multiprocessor systems.

Conditional connectivity introduced by Harary \cite{Harary} can be used to better measure the reliability of multiprocessor systems. If any component of $G-S$ has some property $\mathcal{P}$, where $S$ is a vertex cut of $G$, then $S$ is called a $\mathcal{P}$-$vertex$ $cut$. The $\mathcal{P}$-$conditional$ $connectivity$ of $G$ is defined to be the minimum over all cardinalities of $\mathcal{P}$-vertex cuts. J. F$\grave{a}$brega and M.A. Fiol \cite{F¡®abrega} introduced the $extra$ $connectivity$ of interconnection networks as follows. A vertex set $S\subseteq V(G)$ is called to be an $h$-$extra$ $vertex$ $cut$ if $G-S$ is disconnected and every component of $G-S$ has at least $h+1$ vertices. The $h$-$extra$ $connectivity$ of $G$, denoted by $\kappa_{h}(G)$, is defined as the cardinality of a minimum $h$-extra vertex cut, if exist. An $(h+1)$-extra vertex cut of a graph $G$ is clearly an $h$-extra vertex cut, and thus $\kappa_{h}(G)\leq \kappa_{h+1}(G)$. Extra connectivity is an example of $\mathcal{P}$-conditional connectivity.

It is obvious that $\kappa_{0}(G)=\kappa(G)$ for any graph $G$ that is not a complete graph. In particular, the 1-extra vertex cut is called as the extra vertex cut and the 1-extra connectivity is called as the extra connectivity. The problem of determining the $h$-extra connectivity of numerous networks has received a great deal of attention in recent years. Interested readers may refer to \cite{Balbuena},\cite{Esfahanian.S},\cite{Yang} or others for further details.

The $k$-$ary$ $n$-$cube$ $Q_{n}^{k}$, proposed by Scott and Goodman \cite{Scott}, is one of the most popular interconnection networks. Some properties of the $k$-ary $n$-cube network have been investigated, for example, fault diameter \cite{Day.}, pan-connectivity \cite{Lin} etc. Moreover, many interconnection networks can be viewed as the subclasses of $Q_{n}^{k}$, including the cycle, the torus and the hypercube. The $h$-extra connectivity for $h=1,2$ of $k$-ary $n$-cube are gotten by Hsieh et al. in \cite{Hsieh} for $k\geq4$ and Zhu et al. in \cite{Zhu} for $k=3$. In this paper, we show that the $3$-extra connectivity of the $3$-ary $n$-cube network is $8n-12$ for $n\geq 3$.

Definitions which not been given here are referred to \cite{Bondy} and \cite{Xu}. The remainder of this paper is organized as follows. In Section $2$, the $k$-ary $n$-cube and its properties will be given. Section $3$ discusses the $3$-extra connectivity of the $3$-ary $n$-cube. Section 4 concludes the paper. Last is acknowledgements.

\section{The $k$-ary $n$-cube and its properties}

The $k$-$ary$ $n$-$cube$, denoted by $Q_{n}^{k}$, where $k\geq2$ and $n\geq1$ are integers, is a graph consisting of $k^{n}$ vertices. Each of these vertices has the form $u=u_{n-1}u_{n-2}\cdots u_{0}$ where $u_{i}\in \{0,1,2,\cdots,k-1\}$ for $0\leq i\leq n-1$. Two vertices $u=u_{n-1}u_{n-2}\cdots u_{0}$ and $v=v_{n-1}v_{n-2}\cdots v_{0}$ in $Q_{n}^{k}$ are adjacent if and only if there exists an integer $j$, where $0\leq j\leq n-1$, such that $u_{j}=v_{j}\pm 1(mod$ $k)$, and $u_{i}=v_{i}$ for every $i\in \{0,1,2,\cdots,j-1,j+1,\cdots,n-1\}$. In this case, $(u, v)$ is a $j$-dimensional edge. For clarity of presentation, $``$(mod $k)"$ does not appear in similar expressions in the remainder of the paper. Obviously, $Q_{1}^{k}$ is a cycle of length $k$, $Q_{n}^{2}$ is an $n$-dimensional hypercube, $Q_{2}^{k}$ is a $k\times k$ wrap-around mesh. This study considers $3$-ary $n$-cube, $Q_{2}^{3}$ and $Q_{3}^{3}$ are illustrated in Fig.1.

\begin{figure}
\begin{center}
\includegraphics[height=5cm,width=12cm]{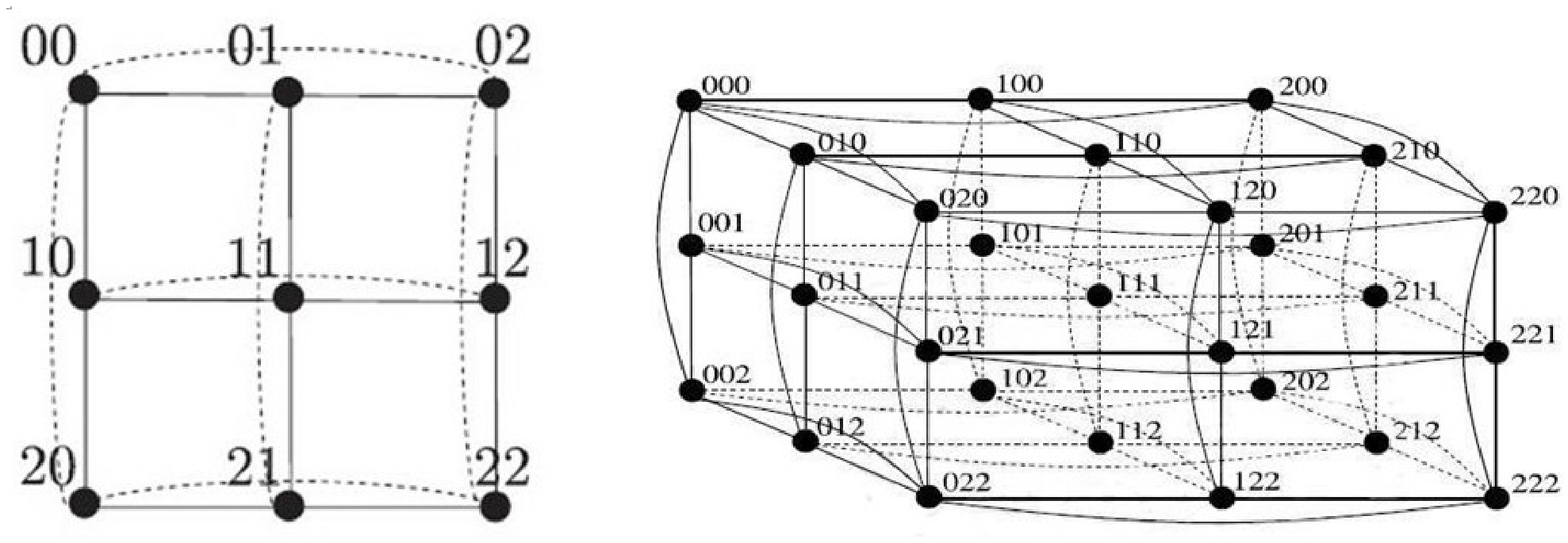}
\end{center}
\begin{center}
{\bf Fig 1.} The illustration of $Q_{2}^{3}$ and $Q_{3}^{3}$
\end{center}
\end{figure}

It is possible to partition $Q_{n}^{k}$ over $j$-dimension, for a $j\in \{0,1,2,\cdots,n-1\}$, into $k$ disjoint subcubes, denoted by $Q_{n-1}^{k}[0],Q_{n-1}^{k}[1],\cdots,Q_{n-1}^{k}[k-1]$ by deleting all the $j$-dimensional edges from $Q_{n}^{k}$. For convenience, abbreviate these as $Q[0],Q[1],\cdots, Q[k-1]$ if there is no ambiguity. Moreover, $Q[i]$ for $0\leq i\leq k-1$ is isomorphic to $k$-ary $(n-1)$-cube and there are $k^{n-1}$ edges between $Q[i]$ and $Q[i+1]$. For each vertex $u \in V(Q[i])$, the $right$ $neighbor$ (respectively, $left$ $neighbor$) of $u$, denoted by $u_{R}$ (respectively, $u_{L}$), is the $outer$ $neighbor$ of $u$ in $Q[i+1]$ (respectively, $Q[i-1]$).

\medskip
The following useful properties of $Q_{n}^{k}$ which will be used later on can be found in \cite{Bose},\cite{Ghozati},\cite{Scott},\cite{Xu}.

\begin{lem}
 (\cite{Scott})  For $n\geq 1$, $Q_{n}^{k}$ is $n$-regular and has $nk^{n-1}$ edges when $k=2$; $Q_{n}^{k}$ is $2n$-regular and has $nk^{n}$ edges when $k\geq3$.
 \end{lem}

\begin{lem}
(\cite{Ghozati,Xu}) For $n\geq 2$, $\kappa(Q_{n}^{k})=\delta(Q_{n}^{k})=2n$ when $k\geq3$; $\kappa(Q_{n}^{k})=\delta(Q_{n}^{k})=n$ when $k=2$. Moreover, $Q_{n}^{k}$ is super-connected for $n\geq 2$.
\end{lem}

\begin{lem}
(\cite{Ghozati}) For $k\geq2, n\geq 2$, $Q_{n}^{k}$ is vertex transitive and edge transitive.
\end{lem}

\begin{lem}
(\cite{Bose}) For $k\geq3, n\geq 2$, $Q_{n}^{k}$ can be divided into $k$ disjoint subgraphs, each subgraph is isomorphic to $Q_{n-1}^{k}$. The two outer neighbors of every vertex in $Q_{n}^{k}$ are in different subgraphs which contained in $\{Q[i]:0\leq i\leq k-1\}$.
\end{lem}

\section{Main result}
The $h$-extra connectivity for $h=1,2$ of $k$-ary $n$-cube are gotten by Hsieh et al. in \cite{Hsieh} for $k\geq 4$ and Zhu et al. in \cite{Zhu} for $k=3$. Nevertheless, the $h$-extra for $h\geq 3$ connectivity of $3$-ary $n$-cube has not been obtained yet. In this section, the $3$-extra connectivity of the $3$-ary $n$-cube will be determined.

\begin{lem}
(\cite{Zhu})  Any two adjacent vertices in $Q_{n}^{3}$ have exactly one common neighbor for $n\geq 1$; If any two nonadjacent vertices in $Q_{n}^{3}$ have common neighbors, they have exactly two common neighbors for $n\geq 2$.
\end{lem}

\begin{lem}
(\cite{Zhu}) $\kappa_{1}(Q_{n}^{3})=4n-3$ for $n\geq 2$ and $\kappa_{2}(Q_{n}^{3})=6n-7$ for $n\geq 3$.
\end{lem}

\begin{lem}
Suppose that $F\subseteq V(Q_{n}^{3})$ with $|F|\leq 4n-4$ is a vertex cut of $ Q_{n}^{3}$ for $n\geq 2$, then $Q_{n}^{3}-F$ has two components, one of which is a singleton.
\end{lem}

\f {\bf Proof.} This lemma can be proved by using induction on $n$.

For $n=2$, the graph is $Q_{2}^{3}$, shown in Fig.1, in this case, $|F|\leq 4n-4=4$; Since $F$ is a vertex cut of $ Q_{n}^{3}$, then $|F|\geq \kappa(Q_{n}^{3})=2n=4$. Thus $|F|=4$, $F$ is the minimum vertex cut. By Lemma 2.2, $Q_{2}^{3}-F$ has two components, one of which is a singleton.

Assume now $n\geq 3$ and the lemma is true for $Q_{n-1}^{3}$. Recall that $Q[i]$ is $2(n-1)$-regular and is isomorphic to $Q_{n-1}^{3}$ for any $i\in\{0,1,2\}$. Let $F_{i}=F \bigcap V(Q[i])$, so $\sum_{i=0}^{2}|F_{i}|\leq4n-4$. Then at least one of $|F_{0}|$, $|F_{1}|$ and $|F_{2}|$ is strictly less than $2(n-1)$ since $F_{0}\bigcap F_{1}\bigcap F_{2}=\emptyset$ and $|F_{0}|+|F_{1}|+|F_{2}|=|F|\leq 4n-4 <6n-6$ for $n\geq2$. Without loss of generality, suppose $|F_{0}|\leq 2(n-1)-1$. We consider the following cases.

{\bf Case 1.} For any $i\in \{1,2\}$, $|F_{i}|\leq 2(n-1)-1$.

Since $\kappa(Q_{n-1}^{3})=2(n-1)$, then $Q[i]-F_{i}$ is connected. There are $3^{n-1}$ edges between $Q[i]$ and $Q[j]$ for $0\leq i\neq j\leq2$, and $3^{n-1}>2(n-1)-1+2(n-1)-1=4n-6$ for $n\geq2$, then $Q[i]-F_{i}$ is connected to $Q[j]-F_{j}$. By the arbitrary of $i$ and $j$, $Q_{n}^{3}-F$ is connected which is a contradiction.

{\bf Case 2.} $|F_{1}|\leq 2(n-1)-1$ and $2(n-1)\leq|F_{2}|\leq 4(n-1)-4$ (The case of $2(n-1)\leq|F_{1}|\leq 4(n-1)-4$, $|F_{2}|\leq 2(n-1)-1$ is the similar discussion).

By the similar argument to Case 1, $Q[0]-F_{0}$ is connected to $Q[1]-F_{1}$. $Q[i]-F_{i}$ for $i\in\{0,1\}$ belong to a same component of $Q_{n}^{3}-F$, denoted by $C$. If $F_{2}$ is not a vertex cut of $Q[2]$, then $Q[2]-F_{2}$ is connected, there are at least $3^{n-1}-(4n-8)-(2n-3)\geq2$ edges between $Q[i]-F_{i}$ for each $i\in\{0, 1\}$ and $Q[2]-F_{2}$ for $n\geq3$, thus $Q[2]-F_{2}$ is contained in $C$, it implies that $Q_{n}^{3}-F$ is connected which is a contradiction. So $F_{2}$ is a vertex cut of $Q[2]$. By inductive hypothesis, $Q[2]-F_{2}$ has two components, one of which is a singleton. Let $B$ be the largest component of $Q[2]-F_{2}$. We will show that $B$ is connected to $C$. Note that $B$ has at least $3^{n-1}-(4n-8)-1$ vertices, and $2[3^{n-1}-(4n-8)-1]$ different outer neighbors. Since there are at most $4n-4-(2n-2)=2n-2$ vertices in $F-F_{2}$, and $2[3^{n-1}-(4n-8)-1]>2n-2$ for $n\geq3$, there must be an edge between $B$ and $C$, then $B$ is contained in $C$. Thus $Q_{n}^{3}-F$ has two components, one of which is a singleton.

{\bf Case 3.} For some $i\in \{1,2\}$, $|F_{i}|\geq 4n-7$.

Since $2(4n-7)>4n-4$ for $n\geq 3$, there is only one $i\in\{1,2\}$ such that $|F_{i}|\geq 4n-7$. Without loss of generality, let $i=1$. Since $|F|\leq 4n-4$, then $\sum_{j\neq 1}|F_{j}|\leq3$. By the the similar argument as Case 1, $Q[j]-F_{j}$ for $j\in \{0,2\}$ belong to a same component of $Q_{n}^{3}-F$, denoted by $C$. If a component, denoted by $B$, of $Q[1]-F_{1}$ has an edge, then its endpoints have exactly four distinct outer neighbors, so $B$ is contained in $C$. Thus only singletons of $Q[1]-F_{1}$ may not be contained in $C$. Since each singleton of $Q[1]-F_{1}$ has two outer neighbors, which are all different, there can be only
one such singleton. Hence $Q_{n}^{3}-F$ has two components, one of which is a singleton.

{\bf Case 4.} For any $i\in \{1,2\}$, $4n-7\geq |F_{i}|\geq 2(n-1)$.

Since $|F|\leq 4n-4$, then $|F_{0}|\leq 0$, that is $|F_{0}|=0$. So $Q[0]-F_{0}$ is connected, denoted it by $C$. Every vertex of $Q[i]-F_{i}$ has one neighbor outside $Q[1]\bigcup Q[2]$ for each $i\in \{1,2\}$, so any component of $Q[i]-F_{i}$ is connected to $C$. Then $Q_{n}^{3}-F$ is connected which is a contradiction.

The lemma is completed.  \hfill\qed

\begin{lem}
Suppose that $F\subseteq V(Q_{n}^{3})$ with $|F|\leq 6n-8$ is a vertex cut of $ Q_{n}^{3}$ for $n\geq 3$, then $Q_{n}^{3}-F$  either has two components, one of which is a singleton, or an edge; or has three components, two of which are singletons.
\end{lem}

\f {\bf Proof.}  Let $F_{i}=V(Q[i])\bigcap F$ for $i\in\{0,1,2\}$, so $\sum_{i=0}^{2}|F_{i}|=|F|\leq 6n-8$. Since $F_{0}\bigcap F_{1}\bigcap F_{2}=\emptyset$ and $|F_{0}|+|F_{1}|+|F_{2}|=|F|\leq 6n-8 <6n-6$ for $n\geq3$, then at least one of $|F_{0}|$, $|F_{1}|$ and $|F_{2}|$ is strictly less than $2(n-1)$. Without loss of generality, suppose $|F_{0}|\leq 2(n-1)-1$. We will prove the lemma by using induction on $n$.

{\bf Case 1.} We prove the result for $Q_{n}^{3}$ with $n=3$.

The graph is $Q_{3}^{3}$, shown in Fig.1. Let $F\subseteq V(Q_{3}^{3})$ with $|F|\leq 6n-8=10$ be a vertex cut of $ Q_{3}^{3}$. We consider the following subcases.

{\bf Subcase 1.1.} For any $i\in\{1,2\}$, $|F_{i}|\leq3$.

Since $\kappa(Q_{2}^{3})=4$, then $Q[i]-F_{i}$ is connected. There are $3^{2}-|F_{i}|-|F_{j}|\geq 3^{2}-2\times3=3$ edges between $Q[i]-F_{i}$ and $Q[j]-F_{j}$ for $0\leq i\neq j\leq 2$, thus $Q[i]-F_{i}$ is connected to $Q[j]-F_{j}$. By the arbitrary of $i$ and $j$, $Q_{3}^{3}-F$ is connected which is a contradiction.

{\bf Subcase 1.2.} $|F_{1}|\leq3$ and $4=2(n-1)\leq |F_{2}|\leq6(n-1)-8=4$, that is $|F_{2}|=4$ (The case of $|F_{1}|=4$ and $|F_{2}|\leq3$ is the similar discussion).

By the similar argument to Subcase 1.1, $Q[0]-F_{0}$ is connected to $Q[1]-F_{1}$. $Q[i]-F_{i}$ for $i\in\{0,1\}$ belong to a same component of $Q_{n}^{3}-F$, denoted by $C$. If $Q[2]-F_{2}$ is connected, since there are at least $3^{2}-|F_{i}|-|F_{2}|\geq3^{2}-4-3=2$ edges between $Q[i]-F_{i}$ and $Q[2]-F_{2}$ for $i\in\{0,1\}$, then $Q[2]-F_{2}$ is contained in $C$, thus $Q_{3}^{3}-F$ is connected which is a contradiction. Hence $Q[2]-F_{2}$ is disconnected, since $|F_{2}|=\kappa(Q_{2}^{3})=4$, $F_{2}$ is a minimum vertex cut of $Q[2]$, by Lemma 2.2, $Q[2]-F_{2}$ has two components, one of which is a singleton. Let $B$ be the largest component of $Q[2]-F_{2}$. Note that $B$ has $4$ vertices, which has $8$ different outer neighbors. Since there are at most $10-4=6$ vertices in $F-F_{2}$, there must be two edges between $B$ and $C$, then $B$ is contained in $C$. Thus $Q_{3}^{3}-F$ has two components, one of which is a singleton.

{\bf Subcase 1.3.}  For some $i\in \{1,2\}$, $|F_{i}|\geq 6n-13=5$.

Without loss of generality, suppose $|F_{2}|\geq 5$. Since $|F|\leq 10$ and $|F_{0}|\leq3$, then $\sum_{j\neq 2}|F_{j}|\leq5$, $|F_{1}|\leq2$. Since $\kappa(Q_{2}^{3})=4$, then $Q[i]-F_{i}$ is connected for $i\in\{0,1\}$. By the similar argument as Subcase 1.1, $Q[0]-F_{0}$ is connected to $Q[1]-F_{1}$. $Q[i]-F_{i}$ for $i\in\{0,1\}$ belong to a same component of $Q_{n}^{3}-F$, denoted by $C$. Every vertex of $Q[2]-F_{2}$ has two outer neighbors, any component with more than two vertices of $Q[2]-F_{2}$ is contained in $C$. Thus $Q_{3}^{3}-F$  either has two components, one of which is a singleton, or an edge; or has three components, two of which are singletons.

{\bf Subcase 1.4.} $|F_{1}|\geq4, |F_{2}|\geq 4$.

Since $|F|\leq 10$, then $|F_{0}|=|F|-|F_{1}|-|F_{2}|\leq 2$. So $Q[0]-F_{0}$ is connected. Let $C$ be a component of $Q_{3}^{3}-F$ which contains $Q[0]-F_{0}$. Every vertex of $Q[i]-F_{i}$ has one neighbor outside $Q[1]\bigcup Q[2]$ for $i\in\{1,2\}$, so any component with more than two vertices of $Q[i]-F_{i}$ is contained in $C$. Thus $Q_{3}^{3}-F$ either has two components, one of which is a singleton, or an edge; or has three components, two of which are singletons.

Then the result holds for $n=3$. In what follows, assume that $n\geq 4$ and the result holds for $ Q_{n-1}^{3}$.

{\bf Case 2.} We prove the result for $Q_{n}^{3}$ and $n\geq 4$. We consider the following subcases.

{\bf Subcase 2.1.} For any $i\in\{1,2\}$, $|F_{i}|\leq2(n-1)-1$.

Since $\kappa(Q_{n-1}^{3})=2(n-1)$, then $Q[i]-F_{i}$ is connected. There are $3^{n-1}$ edges between $Q[i]$ and $Q[j]$ for $0\leq i\neq j\leq2$, and $3^{n-1}>2(n-1)-1+2(n-1)-1=4n-6$ for $n\geq3$, then $Q[i]-F_{i}$ is connected to $Q[j]-F_{j}$. By the arbitrary of $i$ and $j$, $Q_{n}^{3}-F$ is connected which is a contradiction.

{\bf Subcase 2.2.} $|F_{1}|\leq 2(n-1)-1$ and $2(n-1)\leq|F_{2}|\leq 6n-14$ (The case of $2(n-1)\leq|F_{1}|\leq 6n-14$ and $|F_{2}|\leq 2(n-1)-1$ is the similar discussion).

By the similar argument as Subcase 2.1, $Q[0]-F_{0}$ is connected to $Q[1]-F_{1}$. The component which contains $Q[i]-F_{i}$ for $i\in\{0,1\}$ of $Q_{n}^{3}-F$ is denoted by $C$. If $Q[2]-F_{2}$ is connected, there are at least $3^{n-1}-(6n-14)-(2n-3)\geq12$ edges between $Q[i]-F_{i}$ (for $i\in\{0,1\}$) and $Q[2]-F_{2}$ for $n\geq4$, thus $Q[2]-F_{2}$ is connected to $C$. It implies that $Q_{n}^{3}-F$ is connected which is a contradiction. Hence $F_{2}$ is a vertex cut of $Q[2]$. By inductive hypothesis, there are at most three components in $Q[2]-F_{2}$, with two of them having at most two vertices in total. Let $D$ be the largest component of $Q[2]-F_{2}$. $D$ has at least $3^{n-1}-(6n-14)-2$ vertices, by Lemma 2.4, $D$ has at least $2[3^{n-1}-(6n-14)-2]$ distinct outer neighbors. Since $|F_{0}|+|F_{1}|=|F|-|F_{2}|\leq6n-8-(2n-2)=4n-6$ and $2[3^{n-1}-(6n-14)-2]>4n-6$ for $n\geq4$, then there exists at least one edge between $D$ and $C$. Hence $D$ is contained in $C$. Thus the smallest component at most contains two vertices, the result is proved in this case.

{\bf Subcase 2.3.} For some $i\in \{1,2\}$, $|F_{i}|\geq 6n-13$.

Without loss of generality, suppose $|F_{2}|\geq 6n-13$. Then $\sum_{j\neq 2}|F_{j}|=|F|-|F_{2}|\leq 5\leq2(n-1)-1$ for $n\geq4$. Since $\kappa(Q_{n-1}^{3})=2(n-1)\geq6$ for $n\geq4$, then $Q[i]-F_{i}$ is connected for each $i\in\{0,1\}$. By the similar argument as Subcase 2.1, $Q[0]-F_{0}$ is connected to $Q[1]-F_{1}$. Let $C$ be the component of $Q_{n}^{3}-F$ which contains $Q[i]-F_{i}$ for $i\in\{0,1\}$. Every vertex has two outer neighbors, any component of $Q[2]-F_{2}$ with more than two vertices is contained in $C$. Thus $Q_{n}^{3}-F$ either has two components, one of which is a singleton, or an edge; or has three components, two of which are singletons.

Next we only need to consider the case that $F_{i}$ is greater than $2(n-1)-1$ for each $i\in\{1,2\}$.

{\bf Subcase 2.4.} $|F_{1}|\geq|F_{2}|\geq 2(n-1)$ (The case of $|F_{2}|\geq|F_{1}|\geq 2(n-1)$ is the similar discussion).

Clearly, $|F_{2}|\leq|F_{1}|\leq 6n-8-2(n-1)=4n-6$. If $|F_{1}|=4n-6$, then $|F_{2}|=2n-2$, $|F_{0}|=0$. So $Q[0]-F_{0}$ is connected, denoted it by $C$. Every vertex of $Q[i]-F_{i}$ has one neighbor outside $Q[1]\bigcup Q[2]$ for each $i\in\{1,2\}$, so any component of $Q[i]-F_{i}$ is contained in $C$. Then $Q_{n}^{3}-F$ is connected which is a contradiction.

Now we consider $|F_{1}|\leq 4n-7$. First consider $|F_{1}|=4n-7$, if $|F_{2}|=2n-1$, we have done by the previous argument. So the left case is $|F_{2}|=2n-2$, $|F_{0}|=1$. Thus $Q[0]-F_{0}$ is connected, denoted it by $C$. Since every vertex in $Q[1]$ and $Q[2]$ has an outer neighbors in $Q[0]$ and $|F_{0}|=1$, then at most one vertex can be disconnected from $C$ in $Q_{n}^{3}-F$, hence $Q_{n}^{3}-F$ has two components, one of which is a singleton.

Finally, we consider those cases where $|F_{2}|\leq|F_{1}|\leq 4n-8=4(n-1)-4$. This case is divided into three subcases.

{\bf Subcase 2.4.1.} Both $Q[1]-F_{1}$ and $Q[2]-F_{2}$ are connected.

 In this case, there are at least $3^{n-1}-|F_{i}|-|F_{0}|\geq 3^{n-1}-(4n-8)-(2n-3)\geq14$ edges between $Q[0]-F_{0}$ and $Q[i]-F_{i}$ for each $i\in \{1,2\}$ and $n\geq4$, thus $Q[i]-F_{i}$ is connected to $Q[0]-F_{0}$. Hence $Q_{n}^{3}-F$ is connected which is a contradiction.

{\bf Subcase 2.4.2.} Only one of $Q[1]-F_{1}$ and $Q[2]-F_{2}$ is connected.

Without loss of generality, assume that $Q[1]-F_{1}$ is connected and $Q[2]-F_{2}$ is disconnected. By the similar argument as Subcase 2.4.1, $Q[1]-F_{1}$ is connected to $Q[0]-F_{0}$. Let $C$ be the component of $Q_{n}^{3}-F$ which contains $Q[0]-F_{0}$ and $Q[1]-F_{1}$. Since $Q[2]-F_{2}$ is disconnected and $|F_{2}|\leq 4n-8=4(n-1)-4$, by Lemma 3.3, $Q[2]-F_{2}$ has two components, one of which is a singleton. Let $D$ be the largest component of $Q[2]-F_{2}$. Note that $D$ has at least $3^{n-1}-(4n-8)-1$ vertices, and has at least $3^{n-1}-(4n-8)-1$ neighbors in $Q[0]$, since $3^{n-1}-(4n-8)-1>2(n-1)-1\geq|F_{0}|$ for $n\geq 4$, thus $D$ is contained in $C$. Hence $Q_{n}^{3}-F$ has two components, one of which is a singleton.

{\bf Subcase 2.4.3.} Both $Q[1]-F_{1}$ and $Q[2]-F_{2}$ are disconnected.

By Lemma 3.3, $Q[i]-F_{i}$ for each $i\in\{1,2\}$ has two components, one of which is a singleton, denoted by $x_{i}$. Since $|F_{0}|\leq2(n-1)-1$, then $Q[0]-F_{0}$ is connected. Let $B_{i}$ be the largest component of $Q[i]-F_{i}$ for each $i\in\{1,2\}$. Note that $B_{i}$ has at least $3^{n-1}-(4n-8)-1$ vertices, and has at least $3^{n-1}-(4n-8)-1$ outer neighbors in $Q[0]$, since $3^{n-1}-(4n-8)-1>2(n-1)-1$ for $n\geq4$, thus $B_{i}$ is connected to $Q[0]-F_{0}$. Let $C$ be the component of $Q_{n}^{3}-F$ which contains $B_{i}$ and $Q[0]-F_{0}$.

If both $x_{1}$ and $x_{2}$ are contained in $C$, then $Q_{n}^{3}-F$ is connected which is a contradiction. If only one of $x_{i}$ is contained in $C$, then $Q_{n}^{3}-F$ has two components, one of which is a singleton. Besides, the two singletons in $Q[1]-F_{1}$ and $Q[2]-F_{2}$ may either remain singleton components in $Q_{n}^{3}-F$ or they could belong to one component of $Q_{n}^{3}-F$, forming a $K_{2}$. Hence $Q_{n}^{3}-F$ either has two components, one of which is an edge; or has three components, two of which are singletons. The result holds in this case.

The proof of the lemma is finished. \hfill\qed

\begin{theorem}
 For $n\geq 3$, $\kappa_{3}(Q_{n}^{3})\leq 8n-12$.
\end{theorem}

\f {\bf Proof.} Let $u=(0,0,0,0,\cdots,0)$, $v=(0,1,0,0,\cdots,0)$, $w=(0,1,1,0,\cdots,0)$ and $t=(0,0,1,0,\cdots,0)$ be four vertices in $Q_{n}^{3}$, $P_{4}=uvwt \in Q_{n}^{3}$ be a path of length three. Let $F=N(P_{4})$, obviously, $Q_{n}^{3}-F$ is disconnected. Note that $(u,v,w,t,u)$ is a cycle of length four. By Lemma 3.1 and the structure of $Q_{n}^{3}$, $u$ has a neighbor set $X_{1}$ with the order $2n-2$ in $V(Q_{n}^{3})-V(P_{4})$; Since $u$ and $v$ have one common neighbor $x_{1}=(0,2,0,0,\cdots,0)$, $v$ has a neighbor set $X_{2}$ with the order $2n-2-1=2n-3$ in $V(Q_{n}^{3})-V(P_{4})-X_{1}$; Since $w$ and $u$ have two common neighbors $v$ and $t$, $w$ and $v$ have one common neighbor $x_{2}=(0,1,2,0,\cdots,0)$, $w$ has a neighbor set $X_{3}$ with the order $2n-2-1=2n-3$ in $V(Q_{n}^{3})-V(P_{4})-X_{1}-X_{2}$; Since $t$ and $u$ have one common neighbor $x_{3}=(0,0,2,0,\cdots,0)$, $t$ and $v$ have two common neighbors $w$ and $u$, $t$ and $w$ have one common neighbor $x_{4}=(0,2,1,0,\cdots,0)$, $t$ has a neighbor set $X_{4}$ with the order $2n-2-1-1=2n-4$ in $V(Q_{n}^{3})-V(P_{4})-X_{1}-X_{2}-X_{3}$. Thus  $|F|=|X_{1}|+|X_{2}|+|X_{3}|+|X_{4}|=(2n-2)+(2n-3)+(2n-3)+(2n-4)=8n-12$. We will show that $F$ is a $3$-extra vertex cut of $Q_{n}^{3}$ for $n\geq 3$.

For $n=3$, from Fig.1, it is easy to see that $F$ is a $3$-extra vertex cut of $Q_{3}^{3}$. We assume that $n\geq 4$ in the following. Recall that $N[P_{4}]=N(P_{4})\bigcup P_{4}$, we will prove that $Q_{n}^{3}- N[P_{4}]$ is connected for $n\geq 4$.

Without loss of generosity, we partition $Q_{n}^{k}$ over $0$-dimension. Let $F_{i}=N(P_{4})\bigcap Q[i]$, where $i\in\{0,1,2\}$. Note that $P_{4}=uvwt\in Q[0]$, the two outer neighbors of every vertex in $P_{4}$ are in different subgraph $Q[j]$, $j\neq i$, thus $|F_{k}|=4$, $k=1,2$. By Lemma 2.2, $\kappa(Q_{n-1}^{3})=2(n-1)\geq 6>4$, then $Q[k]-F_{k}$ $(k=1,2)$ is connected. Since there are $3^{n-1}\geq 9$ edges between $Q[i]$ and $Q[j]$ for $0\leq i\neq j\leq2$ and $n\geq 3$. Thus $Q_{n}^{3}-Q[0]-N[P_{4}]$ is connected, denoted by $C$.

Now we consider $Q[0]-N[P_{4}]$, for any $ x\in Q[0]-N[P_{4}]$, $x$ has two outer neighbors $x_{L}$ and $x_{R}$, obviously, $x_{L}$($x_{R}$) is not in $N[P_{4}]\bigcap(Q_{n}^{3}-Q[0])$. Hence $x$ is connected to $C$. By the arbitrary of $x$, $C\bigcup(Q[0]-N[P_{4}])=Q_{n}^{3}-N[P_{4}]$ is connected.

Thus $Q_{n}^{3}-F$ has two components, $Q_{n}^{3}-N[P_{4}]$ and $P_{4}$. Then $F$ is a $3$-extra vertex cut of $Q_{n}^{3}$ for $n\geq 3$, thus $\kappa_{3}(Q_{n}^{3})\leq |F|=8n-12$. The theorem is completed. \hfill\qed

\medskip

In the following, suppose $F\subseteq V(Q_{n}^{3})$ is a faulty vertex set of $ Q_{n}^{3}$. For convenience, let $F_{i}=F \bigcap V(Q[i])$, $I=\{i|$ $Q[i]-F_{i}$ is disconnected for $i\in\{0,1,2\}\}$, $J=\{0,1,2\}\backslash I$. $F_{I}=\bigcup_{i\in I}F_{i}$, $F_{J}=\bigcup_{j\in J}F_{j}$, $ Q[I]=\bigcup_{i\in I}Q[i]$, $ Q[J]=\bigcup_{j\in J}Q[j]$.

\begin{lem}
Let $F\subseteq V(Q_{n}^{3})$ with $|F|\leq 8n-13$ be a faulty vertex set of $Q_{n}^{3}$. Then, $Q[J]-F_{J}$ is connected when $|I|\leq 2$ and $n\geq4$. Furthermore, let $H$ be a component of $Q_{n}^{3}-F$ and $H\bigcap(Q[J]-F_{J})=\emptyset$, then $N_{Q[I]}(H)\subseteq F_{I}$, $N_{Q[J]}(H)\subseteq F_{J}$.
\end{lem}

\f {\bf Proof.} Clearly, $|I|\leq 3$, $|F_{i}|\geq 2n-2$ for any $i\in I$. By the definition of $J$, for any $j\in J$, $Q[j]-F_{j}$ is connected. We consider the following three cases.

{\bf Case 1.} $|I|=0$.

For any $j\in J$, $Q[j]-F_{j}$ is connected. Since $3^{n-1}-(8n-13)\geq 8$ for $n\geq4$, there must be an edge between $Q[j]-F_{j}$ and $Q[k]-F_{k}$ for $j\neq k$ and $j,k\in \{0,1,2\}$. $Q[J]-F_{J}$ is connected for $n\geq4$.

{\bf Case 2.} $|I|=1$.

Without loss of generality, we assume that $I=\{0\}$, then $|F_{0}|\geq \kappa(Q_{n-1}^{3})= 2(n-1)$, both $Q[1]-F_{1}$ and $Q[2]-F_{2}$ are connected. There are at least $3^{n-1}-(|F|-|F_{0}|)\geq 3^{n-1}-[8n-13-(2n-2)]\geq3^{3}-(19-6)=14$ edges between $Q[1]-F_{1}$ and $Q[2]-F_{2}$ for $n\geq4$, thus $Q[J]-F_{J}$ is connected.

{\bf Case 3.} $|I|=2$.

Without loss of generality, we assume that $I=\{0,1\}$. Then $Q[J]-F_{J}=Q[2]-F_{2}$ is also connected.

Hence $Q[J]-F_{J}$ is connected for $|I|\leq 2$ and $n\geq4$.

\medskip
Suppose there exists a vertex $u\in N_{Q[I]}(H)$, $u\notin F_{I}$. Then $u$ is connected to $H$, hence $u$ belongs to $H$, which leads to a contradiction. Thus $N_{Q[I]}(H)\subseteq F_{I}$. If there exists a vertex $v\in N_{Q[J]}(H)$, $v\notin F_{J}$, then $v$ is connected to $H$, $v$ is contained in $H$, then $H\bigcap (Q[J]-F_{J})=\{v\}$, which is contradict to $H\bigcap (Q(J)-F_J)=\emptyset$. Hence, $N_{Q[J]}(H)\subseteq F_{J}$. The lemma is completed.\hfill\qed

\begin{theorem}
 Suppose that $F\subseteq V(Q_{n}^{3})$ with $|F|\leq 8n-13$ is a vertex cut of $ Q_{n}^{3}$ for $n\geq 3$, then $Q_{n}^{3}-F$ has one of the following conditions:

 $(1)$ two components, one of which is a singleton or an edge or a 2-path or a 3-cycle.

 $(2)$ three components, two of which are singletons.

 $(3)$ three components, two of which are a singleton and an edge, respectively.

 $(4)$ four components, three of which are singletons.

 \end{theorem}

\f {\bf Proof.} This theorem can be proved by using induction on $n$.

{\bf Case 1.}
Suppose that $n=3$ and $F\subseteq V(Q_{3}^{3})$ with $|F|\leq 8n-13=11$, is a vertex cut of $ Q_{3}^{3}$. We consider the following four subcases.

{\bf Subcase 1.1.}  $|I|=0$.

Since  $J=\{0,1,2\}\backslash I=\{0,1,2\}$.

If for any $j\in J$, $|F_{j}|<\kappa(Q_{n-1}^{3})=2(n-1)=4$. Since $3^{3-1}-|F_{j}|-|F_{k}|\geq 3^{3-1}-4(3-1)=1$, there must be an edge between $Q[j]-F_{j}$ and $Q[k]-F_{k}$ for $j\neq k$ and $j,k\in \{0,1,2\}$. In this case $Q[J]-F_{J}$ is connected which is a contradiction.

If there exists only one $j\in J$, $|F_{j}|\geq\kappa(Q_{n-1}^{3})=2(n-1)=4$, suppose $j=0$. $Q[1]-F_{1}$ is connected to $Q[2]-F_{2}$. Since $|V(Q[0])|=3^{2}=9$, then $4\leq|F_{0}|\leq9$. If $Q[0]-F_{0}$ is empty, then $\sum_{j=1}^{2}|F_{j}|\leq 2$, there are at least $9-2=7$ edges between $Q[1]-F_{1}$ and $Q[2]-F_{2}$, thus $Q_{3}^{3}-F$ is connected. If $Q[0]-F_{0}$ has one vertex, then $\sum_{j=1}^{2}|F_{j}|\leq 3$. By Lemma 2.4, $Q_{3}^{3}-F$ is connected or has two components, one of which is a singleton. If $Q[0]-F_{0}$ has two vertices, then $\sum_{j=1}^{2}|F_{j}|\leq 4$. In this case, $Q_{3}^{3}-F$ is connected or has two components, one of which is an edge. If $Q[0]-F_{0}$ has three vertices, then $\sum_{j=1}^{2}|F_{j}|\leq 5$. The three vertices have six outer neighbors, $Q_{3}^{3}-F$ is connected in this case. If $Q[0]-F_{0}$ has four or five vertices, $Q_{3}^{3}-F$ is connected by the similar reason.

If there exists only two $j\in J$, $|F_{j}|\geq\kappa(Q_{n-1}^{3})=2(n-1)=4$, suppose $j\in\{0,1\}$. Then $|F_{2}|\leq 3$. Clearly, $|F_{0}|\leq|F|-|F_{1}|\leq7$, $|F_{1}|\leq|F|-|F_{0}|\leq7$. Suppose $|F_{0}|\leq|F_{1}|$. If $|F_{1}|=7$, then $|F_{0}|=4$, $|F_{2}|=0$, every vertex of $Q[j]-F[j]$ for $j\in\{0,1\}$ has an outer neighbor in $Q[2]$, then $Q_{3}^{3}-F$ is connected. Now we consider $|F_{1}|=6$, if $|F_{0}|=5$, as the similar reason, $Q_{3}^{3}-F$ is connected. The left case is $|F_{0}|=4$, $|F_{2}|\leq1$, note that $|V(Q[0]-F[0])|=5$ and $|V(Q[1]-F[1])|=3$, every vertex of $Q[j]-F[j]$ for each $j\in\{0,1\}$ has an outer neighbor in $Q[2]$. Thus there is an edge between $Q[j]-F[j]$ and  $Q[2]-F[2]$, then $Q_{3}^{3}-F$ is connected. Next consider $|F_{1}|=5$, if $|F_{0}|=5$, then $|F_{2}|\leq 1$, $Q_{3}^{3}-F$ is connected; if $|F_{0}|=4$, then $|F_{2}|\leq 2$, $Q_{3}^{3}-F$ is also connected.

If for any $j\in J$, $|F_{j}|\geq\kappa(Q_{n-1}^{3})=2(n-1)=4$, then $|F|\geq12$, this is contradict to $|F|\leq 8n-13=11$.

In summary, $Q_{3}^{3}-F$ has two components, one of which is a singleton or an edge. The result holds.

{\bf Subcase 1.2.}  $|I|=1$.

Without loss of generality, we assume that $I=\{0\}$. Then $Q[0]-F_{0}$ is disconnected. By Lemma 2.2, $|F_{0}|\geq 2n-2=4$. Thus $|F|-|F_{0}|\leq 11-4=7$. There are at least $3^{3-1}-(|F|-|F_{0}|)\geq 3^{2}-(11-4)\geq2$ edges between $Q[1]-F_{1}$ and $Q[2]-F_{2}$, then $Q[J]-F_{J}$ is connected.

Suppose $W$ is the union of all components of $Q_{3}^{3}-F$ and has no vertices in $Q[J]-F_{J}$. Since $Q_{3}^{3}-F$ is disconnected, then $W$ exists. Every vertex in $W$ has two outer neighbors, so $2|W|\leq |F|-|F_{0}|\leq7$, that means $|W|\leq 3$, the result holds.

{\bf Subcase 1.3.} $|I|=2$.

Without loss of generality, we assume that $I=\{0,1\}$. Then $Q[0]-F_{0}$ and $Q[1]-F_{1}$ are disconnected, $Q[2]-F_{2}$ is connected. By Lemma 2.2, $|F_{0}|\geq 2n-2=4$ and $|F_{1}|\geq 2n-2=4$, so  $|F_{2}|=|F|-|F_{0}|-|F_{1}|\leq11-4-4=3$. Suppose $W$ is the union of all components of $Q_{3}^{3}-F$ and has no vertices in $Q[2]-F_{2}$. Since $Q_{3}^{3}-F$ is disconnected, then $W$ exists. Every vertex in $W$ has one outer neighbor in $Q[2]$, so $|W|\leq |F_{2}|\leq 3$. The desired result.

{\bf Subcase 1.4.}  $|I|=3$.

In this case, $I=\{0,1,2\}$. For any $i\in I$, $|F_{i}|\geq 2n-2=4$, then $|F|=|F_{0}|+|F_{1}|+|F_{2}|\geq 12$, this is contradict to $|F|\leq 11$.

In summary, we have proved the result holds for $ Q_{3}^{3}$.

\medskip
In what follows, assume that $n\geq 4$ and the result holds for $ Q_{n-1}^{3}$.

{\bf Case 2.} We prove the result for $ Q_{n}^{3}$ and $n\geq 4$. We divide the proof into the following two subcases.

{\bf Subase 2.1.} For any $i\in \{0,1,2\}$, $|F_{i}|\leq 6n-14$.

Recall that $|F|\leq 8n-13$, by Lemma 3.6, $Q[J]-F_{J}$ is connected when $|I|\leq 2$. We consider the following four subcases.

{\bf Subcase 2.1.1.}  $|I|=0$.

Since $J=\{0,1,2\}\backslash I=\{0,1,2\}$, by Lemma 3.6, $ Q[J]-F_{J}=Q_{n}^{3}-F$ is connected which is a contradiction.

{\bf Subcase 2.1.2.}  $|I|=1$.

Without loss of generality, we assume that $I=\{0\}$. By Lemma 3.6, $Q[J]-F_{J}$ is connected. Since $Q[0]-F_{0}$ is disconnected, and $|F_{i}|\leq 6n-14=6(n-1)-8$, by Lemma 3.4, $Q[0]-F_{0}$ either has two components, one of which is a singleton or an edge, denoted by $X_{0}$; or has three components, two of which are singletons, denoted by $X_{0}=\{u,v\}$. Let $B$ be the largest component of $Q[0]-F_{0}$. Next we show $B$ is connected to $Q[J]-F_{J}$. Note that $B$ has at least $3^{n-1}-(6n-14)-2$ vertices, and has at least $2[3^{n-1}-(6n-14)-2]$ outer neighbors in $Q[J]$. Since $2[3^{n-1}-(6n-14)-2]> 8n-13-(2n-2)=6n-11$ for $n\geq4$, then $B$ is connected to $Q[J]-F_{J}$. Thus $Q_{n}^{3}-F$ either has two components, one of which is a singleton or an edge; or has three components, two of which are singletons. The result holds.

{\bf Subcase 2.1.3.} $|I|=2$.

Without loss of generality, we assume that $I=\{0,1\}$. By Lemma 3.6, $Q[2]-F_{2}$ is connected. For $i\in I$, since $|F_{i}|\leq 6n-14=6(n-1)-8$, by Lemma 3.4, $Q[i]-F_{i}$ either has two components, one of which is a singleton or an edge, denoted by $X_{i}$; or has three components, two of which are singletons, denoted by $X_{i}=\{u_{i},v_{i}\}$.  Let $B_{i}$ be the largest component of $Q[i]-F_{i}$, then $X_{i}=Q[i]-F_{i}-B_{i}$. We claim that $B_{i}$ is connected to $Q[2]-F_{2}$. In fact, $B_{i}$ has at least $3^{n-1}-(6n-14)-2$ vertices, and $3^{n-1}-(6n-14)-2$ outer neighbors in $Q[2]$, since $|F_{2}|=|F|-|F_{0}|-|F_{1}|\leq 8n-13-2(2n-2)=4n-9$ and $3^{n-1}-(6n-14)-2 >4n-9$ for $n\geq4$, then $B_{i}$ is connected to $Q[2]-F_{2}$. Let $C$ be the component of $Q_{n}^{3}-F$ which contains $B_{i}$ and $Q[2]-F_{2}$. Next we consider the following three subcases.

{\bf Subcase 2.1.3a.} Both $Q[0]-F_{0}$ and $Q[1]-F_{1}$ have two components, one of which is a singleton or an edge.

If both $Q[0]-F_{0}$ and $Q[1]-F_{1}$ have two components, one of which is a singleton. By the similar argument as Subcase 2.4.3 of Lemma 3.4, $Q_{n}^{3}-F$ either has three components, two of which are singletons; or has two components, one of which is a singleton or an edge. The result holds.

If only one of $Q[0]-F_{0}$ and $Q[1]-F_{1}$ has two components, one of which is a singleton. Without loss of generality, assume that $Q[0]-F_{0}$ has two components, and one of which is a singleton which is denoted by $x_{0}$ and $Q[1]-F_{1}$ has two components, and one of which is an edge which is denoted by $X_{1}=u_{1}v_{1}$. If both $x_{0}$ and $X_{1}$ are contained in $C$, then $Q_{n}^{3}-F$ is connected which is a contradiction. If $x_{0}$ is contained in $C$, $X_{1}$ is not contained in $C$, then $Q_{n}^{3}-F$ has two components, one of which is an edge. If $x_{0}$ is not contained in $C$ and $X_1$ is contained in $C$, then $Q_{n}^{3}-F$ has two components, one of which is a singleton. Besides, the singleton $x_{0}$ in $Q[0]-F_{0}$ and the isolated edge $X_{1}$ in $Q[1]-F_{1}$ may either remain singleton and isolated edge in $Q_{n}^{3}-F$; or they could belong to one component of $Q_{n}^{3}-F$, forming a $2$-path. Thus $Q_{n}^{3}-F$ either has three components, two of which are a singleton and an edge, respectively; or has two components, one of which is a $2$-path. The result holds in this case.

If both $Q[0]-F_{0}$ and $Q[1]-F_{1}$ have two components, one of which is an edge, denoted by $X_{i}$ for $i\in I$. By Lemma 3.3, $|F_{i}|\geq 4(n-1)-3=4n-7$, $|F_{2}|=|F|-|F_{0}|-|F_{1}|\leq 8n-13-2(4n-7)=1$. Every vertex in $X_{i}$ has one outer neighbor in $Q[2]$. By Lemma 3.6, $N_{Q[2]}(X_{i})\subseteq F_{2}$, then $2=|N_{Q[2]}(X_{i})|\leq|F_{2}|\leq 1$ which is a contradiction.

{\bf Subcase 2.1.3b.} Only one of $Q[0]-F_{0}$ and $Q[1]-F_{1}$ has two components.

Without loss of generality, assume that $Q[0]-F_{0}$ has two components and $Q[1]-F_{1}$ has three components.

If $Q[0]-F_{0}$ has two components, one of which is a singleton, denoted by $x_{0}$. $Q[1]-F_{1}$ has three components, two of which are singletons, denoted by $u_{1},v_{1}$. If $x_{0}$, $u_{1}$ and $v_{1}$ are contained in $C$, then $Q_{n}^{3}-F$ is connected which is a contradiction. If $x_{0}$ is contained in $C$, only one of $u_{1}$ and $v_{1}$ is contained in $C$, then $Q_{n}^{3}-F$ has two components, one of which is a singleton. If $x_{0}$ is contained in $C$, both $u_{1}$ and $v_{1}$ are not contained in $C$, then $Q_{n}^{3}-F$ has three components, two of which are singletons. If $x_{0}$ is not contained in $C$, both $u_{1}$ and $v_{1}$ are contained in $C$, $Q_{n}^{3}-F$ has two components, one of which is a singleton. If $x_{0}$ is not contained in $C$, only one of $u_{1}$ and $v_{1}$ is contained $C$. In this case, $Q_{n}^{3}-F$ has two components, one of which is an edge; or $Q_{n}^{3}-F$ has three components, two of which are singletons. Besides, the singletons in $Q[1]-F_{1}$ and $Q[0]-F_{0}$  may remain singleton components in $Q_{n}^{3}-F$; or they could belong to two components of $Q_{n}^{3}-F$, $K_{2}$ and a singleton. Thus $Q_{n}^{3}-F$ either has four components, three of which are singletons; or has three components, two of which are a singleton and an edge, respectively. The result holds.

If $Q[0]-F_{0}$ have two components, one of which is an edge, denoted by $X_{0}$. By Lemma 3.3, $|F_{0}|\geq 4(n-1)-3=4n-7$ and $|F_{1}|\geq 4(n-1)-3=4n-7$, by the similar argument as Subcase 2.1.3a, $2=|N_{Q[2]}(X_{0})|\leq|F_{2}|\leq 1$ which is a contradiction.

{\bf Subcase 2.1.3c.} Both $Q[0]-F_{0}$ and $Q[1]-F_{1}$ have three components, two of which are singletons, denoted by $X_{i}=\{u_{i},v_{i}\}$ for $i\in I$.

By Lemma 3.3, $|F_{0}|\geq 4(n-1)-3=4n-7$ and $|F_{1}|\geq 4(n-1)-3=4n-7$. Then $|F_{2}|=|F|-|F_{0}|-|F_{1}|\leq 8n-13-2(4n-7)=1$. By Lemma 3.6, $N_{Q[2]}(u_{i})\subseteq F_{2}$, and $N_{Q[2]}(v_{i})\subseteq F_{2}$. Then $2=|N_{Q[2]}(X_{i})|\leq|F_{2}|\leq 1$ which is a contradiction.

\medskip
{\bf Subcase 2.1.4.}  $|I|=3$.

We have $I=\{0,1,2\}$, by Lemma 2.2, for any $i\in I$, $|F_{i}|\geq2(n-1)=2n-2$. Since $|F|\leq 8n-13$, then $|F_{i}|\leq 8n-13-2(2n-2)=4n-9\leq 4(n-1)-4$. By Lemma 3.3, $Q[i]-F_{i}$ has two components, one of which is a singleton $x_{i}$. Let $B_{i}$ be the largest component of $Q[i]-F_{i}$ for $i\in I$. Note that $B_{i}$ has at least $3^{n-1}-(4n-9)-1$ vertices, and has at least $3^{n-1}-(4n-9)-1$ neighbors in $Q[k]$ for $0\leq k\neq i\leq2$. $B_{i}$ is connected to $B_{k}$ since $|F_{k}|\leq 4n-9$ and $3^{n-1}-(4n-9)-1>4n-9$ for $n\geq4$. Let $C$ be the component of $Q_{n}^{3}-F$ which contains $B_{i}$ for $i\in I$. We consider the following four subcases.

{\bf Subcase 2.1.4a.} For any $i\in I$, $x_{i}$ is contained in $C$.

In this case, $Q_{n}^{3}-F$ is connected which is a contradiction.

{\bf Subcase 2.1.4b.} There exists only two $i\in I$, such that $x_{i}$ is contained in $C$.

In this case, $Q_{n}^{3}-F$ has two components, one of which is a singleton.

{\bf Subcase 2.1.4c.} Only one $i\in I$, such that $x_{i}$ is contained in $C$.

In this case, $Q_{n}^{3}-F$ either has three components, two of which are singletons; or has two components, one of which is an edge. The result holds.

{\bf Subcase 2.1.4d.} For any $i\in I$, $x_{i}$ is not contained in $C$.

In this case, the three singletons in $Q[i]-F_{i}$ may remain singleton components of $Q_{n}^{3}-F$; or they could belong to two components, a singleton and an edge; or they could belong to one component forming a $3$-cycle or a $2$-path. Thus $Q_{n}^{3}-F$ either has four components, three of which are singletons; or has three components, two of which are a singleton and an edge, respectively; or has two components, one of which is a $2$-path, or a $3$-cycle. Then the result holds.

{\bf Subcase 2.2.}  There exists some $i\in \{0,1,2\}$, such that $|F_{i}|\geq 6n-13$.

Without loss of generality, we assume that $|F_{0}|\geq 6n-13$. For $j\in\{1,2\}$, $|F_{j}|\leq 8n-13-(6n-13)=2n<4(n-1)-3=4n-7$ for any $n\geq 4$. We consider the following two subcases.

{\bf Subcase 2.2.1.}  For $j\in\{1,2\}$, $Q[j]-F_{j}$ is connected.

Since there are at least $3^{n-1}-|F_{1}|-|F_{2}|\geq3^{n-1}-2n-2n >11$ edges between $Q[1]-F_{1}$ and $Q[2]-F_{2}$ for $n\geq4$, then $Q[1]-F_{1}$ is connected to $Q[2]-F_{2}$. The component which contains $Q[j]-F_{j}$ for $j\in\{1,2\}$ of $Q_{n}^{3}-F$ is denoted by $C$.

If $|F_{0}|\leq 8(n-1)-13=8n-21$, suppose $Q[0]-F_{0}$ is connected. There are at least $3^{n-1}-(8n-21)-(2n-2)\geq 10$ edges between $Q[j]-F_{j}$ and $Q[0]-F_{0}$ for $n\geq4$, then $Q[0]-F_{0}$ is contained in $C$. It implies that $Q_{n}^{3}-F$ is connected which is a contradiction. Hence $F_{0}$ is a vertex cut of $Q[0]$. By induction on $n$, $Q[0]-F_{0}$ satisfies one of the conditions $(1)$-$(4)$. Let $D$ be the largest component of $Q[0]-F_{0}$. We claim that $D$ is contained in $C$. In fact, $D$ has at least $3^{n-1}-(8n-21)-3$ vertices, and has $3^{n-1}-(8n-21)-3$ outer neighbors in $Q[2]$, since $3^{n-1}-(8n-21)-3 > 8n-13-(2n-2)=6n-11$ for $n\geq4$, then $D$ is contained in $C$. Thus $Q_{n}^{3}-F$ satisfies one of the conditions $(1)$-$(4)$. The result holds in this case.

Suppose $|F_{0}|> 8n-21$, then $|F|-|F_{0}|<8n-13-(8n-21)=8$. Suppose $W$ is the union of the components of $Q_{n}^{3}-F$ and has no vertices in $C$. Since $Q_{n}^{3}-F$ is disconnected, then $W$ exists. By Lemma 3.6, the outer neighbors of $W$ is in $F-F_{0}$. By Lemma 2.4, $2|W|\leq|F|-|F_{0}|<8$, then $|W|<4$. Hence $|W|\leq 3$, the desired result.

{\bf Subcase 2.2.2.}  There exists $j\in \{1,2\}$, $Q[j]-F_{j}$ is disconnected.

Without loss of generality, we assume that $j=1$ and $Q[1]-F_{1}$ is disconnected. By Lemma 2.2, $|F_{1}|\geq \kappa(Q_{n-1}^{3})=2n-2$, $|F_{2}|=|F|-|F_{0}|-|F_{1}|\leq 8n-13-(6n-13)-(2n-2)=2$, then $Q[2]-F_{2}$ is connected. Furthermore, $|F_{0}|\leq 8n-13-(2n-2)=6n-11$, $|F_{1}|\leq8n-13-(6n-13)=2n\leq4(n-1)-4$ for $n\geq4$. By Lemma 3.3, $Q[1]-F_{1}$ has two components, one of which is a singleton $x_{1}$. Let $D$ be the largest component of $Q[1]-F_{1}$, then $D$ is connected to $Q[2]-F_{2}$. In fact, $D$ has at least $3^{n-1}-2n-1$ vertices and $3^{n-1}-2n-1$ neighbors in $Q[2]$, since $|F_{2}|\leq2$ and $3^{n-1}-2n-1-2>0$ for $n\geq4$, then $D$ is connected to $Q[2]-F_{2}$. The component which contains $D$ and $Q[2]-F_{2}$ of $Q_{n}^{3}-F$ is denoted by $C$.

If $Q[0]-F_{0}$ is connected, since $|F_{2}|\leq2$, $|F_{0}|\geq 6n-13$ and $3^{n-1}-(6n-13)\geq 17>|F_{2}|$ for $n\geq4$, there is at least one edge between $Q[0]-F_{0}$ and $Q[2]-F_{2}$. Then $Q[0]-F_{0}$ is connected to $Q[2]-F_{2}$, $Q[0]-F_{0}$ is contained in $C$. Thus $Q_{n}^{3}-F$ has two components, one of which is a singleton. Then the result holds.

If $Q[0]-F_{0}$ is disconnected, suppose $X$ is the union of the components of $Q[0]-F_{0}$ and has no neighbors in $C$. We will show $|X|\leq2$. Suppose by the way of contradiction that $|X|\geq3$. Every vertex in $X$ has one outer neighbor in $Q[2]\bigcap F$, by Lemma 3.6 and $|F_{2}|\leq 2$, thus $|X|\leq2$ which is a contradiction. Let $E$ be the largest component of $Q[0]-F_{0}$, then $E$ is connected to $Q[2]-F_{2}$. In fact, there must be an edge between $Q[0]-F_{0}$ and $Q[2]-F_{2}$ since $|F_{2}|\leq2$ and $3^{n-1}-(2n-2)-2>2\geq|F_{2}|$ for $n\geq4$. Thus $E$ is contained in $C$.

Suppose $W$ is the union of the components of $Q_{n}^{3}-F$ and has no vertices in $C$. Since $Q_{n}^{3}-F$ is disconnected, then $W$ exists. Obviously, $W\subseteq\{x_{1}\}\bigcup X$. Since $|X|\leq 2$, then $|W|\leq 3$. The desired result.

This covers all possibilities and the proof of the theorem is complete.    \hfill\qed

\medskip
The following theorem about the 3-extra connectivity of the $3$-ary $n$-cube network follows from Theorem 3.5 and Theorem 3.7.

\begin{theorem}
 For $n\geq 3$, $\kappa_{3}(Q_{n}^{3})=8n-12$.
\end{theorem}

\f {\bf Proof.} By Theorem 3.7, $\kappa_{3}(Q_{n}^{3})\geq8n-12$ for $n\geq 3$. On the other hand, by Theorem 3.5, $\kappa_{3}(Q_{n}^{3})\leq8n-12$ for $n\geq 3$. Hence, $\kappa_{3}(Q_{n}^{3})=8n-12$ for $n\geq 3$. The proof of the theorem is complete.   \hfill\qed

\section{Conclusion}

In this paper, the 3-extra connectivity of the $3$-ary $n$-cube networks is gotten. The result shows that at least $8n-12$ vertices must be moved to disconnect the $3$-ary $n$-cube for $n\geq3$, provided that the removal of these vertices does not isolate either a singleton, an edge, a $2$-path, or a $3$-cycle. We will further study $3$-extra connectivity of the $k$-ary $n$-cube networks for $k\geq 4$, $h$-extra connectivity of the $k$-ary $n$-cube networks for $h\geq 4$ and $h$-extra connectivity of other interconnection networks. Determining the $h$-extra connectivity of various multiprocessor systems requires further research efforts.

\bigskip

\f {\bf Acknowledgements}

\bigskip
This work was supported by the National Natural Science Foundation of China (11371052, 11271012, 11231008, 11171020).


\begin{thebibliography}{99}
\bibitem{Balbuena}
C. Balbuena, Extraconnectivity of s-geodetic digraphs and graphs, Discrete Mathematics, 195 (1999) 39-52.
\bibitem{Carmona}
C. Balbuena, A. Carmona, J. Fbrega, M.A. Fiol, Extraconnectivity of graphs with large minimum degree and girth, Discrete Mathematics, 167/168 (1997) 85-100.
\bibitem{Bondy}
J.A. Bondy, U.S.R. Murty, Graph Theory with Applications, North-Holland, New York, 1976.
\bibitem{Bose}
B. Bose, B. Broeg, Y. Kwon, Y. Ashir, Lee distance and topological properties of $k$-ary $n$-cubes, IEEE Transactions on Computers, 44 (8) (1995) 1021-1030.
\bibitem{Cheng}
E. Cheng, L. Lipta'k, F. Sala, Linearly many faults in $2$-tree-generated networks, Networks, 55 (2010) 90-98.
\bibitem{Day}
K. Day, The conditional node connectivity of the $k$-ary $n$-cube, Journal of Interconnection Networks, 5 (1) (2004) 13-26.
\bibitem{Day.}
Khaled Day, Abdel Elah, Al-Ayyoub, Fault diameter of $k$-ary $n$-cube networks, IEEE Transactions on Parallel and distributed systems, 8 (9) (1997) 903-907.
\bibitem{Esfahanian}
A.H. Esfahanian, Generalized measures of fault tolerance with application to $n$-cube networks, IEEE Transactions on Computers, 38 (11) (1989) 1586-1591.
\bibitem{Esfahanian.S}
A.H. Esfahanian, S. L. Hakimi, On computing a conditional edge-connectivity of a graph, Information Processing Letters, 27 (4) (1988) 195-199.
\bibitem{F¡®abrega}
J. F$\grave{a}$brega, M.A. Fiol, On the extra connectivity graphs, Discrete Mathematics, 155 (1996) 49-57.
\bibitem{Fbrega}
J. F$\grave{a}$brega, M.A. Fiol, Extraconnectivity of graphs with large girth, Discrete Mathematics, 127 (1994) 163-170.
\bibitem{Ghozati}
S.A. Ghozati, H.C. Wasserman, The $k$-ary $n$-cube network: modeling, topological properties and routing strategies, Computers and Electrical Engineering, 25 (3) (1999) 155-168.
\bibitem{Hao}
R.X. Hao, Y.Q. Feng, J.X. Zhou, Conditional diagnosability of alternating group graphs, IEEE Transactions on Computers, 62 (4) (2013) 827-831.
\bibitem{Hao.}
R.X. Hao, J.X. Zhou, Characterize a kind of fault tolerance of alternating group network, Acta Mathematica Sinica, Chinese Series, 55 (6) (2012) 1055-1066.
\bibitem{Harary}
F. Harary, Conditional connectivity, Networks, 143 (12) (1983) 346-357.
\bibitem{Hsieh}
Sun-Yuan Hsieh, Ying-Hsuan Chang, Extra connectivity of $k$-ary $n$-cube networks, Theoretical Computer Science, 443 (2012) 63-69.
\bibitem{Lin}
S.Y. Hsieh, T.J. Lin, H.L. Huang, Panconnectivity and edge-pancyclicity of $3$-ary $n$-cubes, The Journal of Supercomputing, 42 (2007) 233-255.
\bibitem{LatiL}
S. LatiL, M. Hegde, M. Naraghi-Pour, Conditional connectivity measures for large multiprocessor systems, IEEE Transactions on Computers, 43 (2) (1994) 218-222.
\bibitem{Scott}
S.L. Scott, J.R. Goodman, The impact of pipelined channel on $k$-ary $n$-cube networks. IEEE Transactions on Parallel and Distributed Systems, 5 (1) (1994) 2-16.
\bibitem{Xu}
J.M. Xu, Combination of Network Theory, Science Press, Beijing, 2013.
\bibitem{Yang}
W.H. Yang, J.X. Meng, Extraconnectivity of hypercubes, Applied Mathematics Letters, 22 (6) (2009) 887-891.
\bibitem{Zhu}
Q. Zhu, X.K. Wang, J.J. Ren, Extra connectivity measures of $3$-ary $n$-cubes, Theory of Computing Systems, arxiv.org/pdf/1105.0991v1 [cs.DM] 5 May 2011.

\end{thebibliography}
\end{document}